# The evolution of Liouville von Neumann master equations in the Pechukas–Yukawa framework

Mumnuna A. Qureshi

*Attero Solutions Ltd., Chatham, United Kingdom*

E-mail: mumnunaqureshi@googlemail.com

This paper presents a novel formalism for the out-of-equilibrium dynamics of the density matrix, capable of describing highly entangled many-body interactions. The evolution of quantum states is evaluated via eigenvalue dynamics of a general Hamiltonian system, perturbed by a parametrically evolving variable $\lambda(t)$ that carries the time-dependence. This is achieved using the Pechukas–Yukawa mapping of the evolution of the energy levels governed by their initial conditions on a generalized Calogero–Sutherland model of a 1D classical gas. As such, quantum systems can be described exactly in their entirety from eigenvalue dynamics. Under this description, we provide an improved understanding of the relationship between nonequilibrium quantum phase transitions and decoherence, which has significant impacts on a wide range of applications.

## 1. Introduction

Theoretical description of open quantum many-body systems is a critical area of research with strong connections to modern physics, optical systems, and quantum computing. One of the key challenges to quantum computing is decoherence. A promising alternative approach is adiabatic quantum computing (AQC), encoding the system in an easily achievable ground state of the initial Hamiltonian and evolving under an adiabatic parameter such that the system maps to the ground state of the final Hamiltonian. This corresponds to an optimal solution. The ground state is more robust against decoherence; however, it is not immune to it [1–5]. Due to the persisting obstacle of decoherence, the realization of quantum computers remains a challenging task. Given a physical quantum platform, based on the superposition of states, the system will inevitably be open. It is impossible to perfectly isolate the system from its environment, hence decoherence can be attributed to both quantum phase transitions and noise [1–5]. Open quantum systems have been studied across multiple disciplines through the quantum master equation. For many-body interactions, these have been explored numerically through approximation schemes such as perturbation theory and statistics of spectral quantum trajectories, in the absence of analytical solutions [3, 4, 6, 7]. For higher-order approximations, these methods become computationally expensive in time and memory [3, 4, 6, 7]. In this paper, an analytical solution given by level dynamics offers a more robust description of state dynamics, which can provide useful insights into the mechanisms leading to phase transitions and decoherence [4, 8].

The study of adiabatically evolving systems is important to the development of AQC, which will improve optimization problems. These systems are governed by the Hamiltonian [1–5, 9–16]:

$$H(\lambda(t)) = H_0 + \lambda(t) Z H_b, \qquad (1)$$

where $H_0$ is a complex unperturbed Hamiltonian with an easily achievable nondegenerate ground state, $\lambda$ is an adiabatically evolving parameter, $ZH_b$ is a large bias perturbation term with $Z \gg 1$. There is a remarkable similarity between AQC algorithms and quantum phase transitions in their Hamiltonians governing their dynamics; $H(\lambda(t)) = H_0 + \lambda(t) H_1$, where $H_0$ is the free Hamiltonian and $H_1$ is some perturbation [5]. There exist various models that seek to describe the relations between quantum phase transitions, decoherence, and entanglement. One source of decoherence is noise, interacting with the system, resulting in dissipation in the evolution of states. These may result in variation of occupation numbers; however, our understanding of nonequilibrium dynamics is limited.

Simulating the evolution of a master equation proves more difficult than Hamiltonian dynamics due to the sheer amount of information required for the density matrix. Currently existing models are often reliant on mean field approximations to simplify the system or factorized approximations due to the large amounts of information required [3, 4, 6, 7, 17]; here we offer a seamless description for these properties in the lens of eigenvalue dynamics for a quantum system via the Pechukas–Yukawa formalism, capable of investigating the influence of noise on the evolution of the density matrix. This model is also expressible in Lax formalism, enabling the study of symmetries and conserved quantities in quantum phase transitions; however, this area of research is beyond the scope of this paper.

The Pechukas–Yukawa formalism maps the level dynamics of Eq. (1) to a one-dimensional (1D) classical gas with long-range repulsion [18]. It is especially convenient for AQC, taking λ to be an adiabatically evolving parameter; however, it is not restricted to such systems. In [2], this is used to connect the level dynamics of a system to the quantum states through the evolution of $C(t)$, for a wavefunction expanded in the instantaneous eigenstates $\psi = \sum_n C_n(t) | n \rangle$. Extending this to the description of the density matrix $\rho(t) = C(t) \otimes C(t)$ in this paper, provides insights into the dynamics of occupation numbers. One can determine the effects of anticrossings on the system's evolution and the extent to which the noise affects the population of states and hence determine the probabilities of the system remaining in initial states. In this framework, there is no restriction on the stochastic model describing the influence of the system's interactions with its environment. This provides great insights into and beyond AQC, to the development of non-equilibrium quantum state dynamics, both analytically and within the grasp of experimental testing.

The structure of the paper is as follows: Sec. 1 gives a brief overview of the Pechukas–Yukawa equations, Sec. 3 introduces the stochastic Liouville von Neumann master equation in the Pechukas–Yukawa formalism, exactly describing quantum states without approximation. Section 4 provides numerical representations highlighting the use of the stochastic master equation for the two-qubit Ising model under the influences of noise, and Sec. 5 presents discussions and conclusions.

## 2. The Pechukas model and the evolution of eigenstate coefficients

The Pechukas equations model a classical fictitious gas moving in 1D with parametric evolution in time, well-suited though not restricted to adiabatic systems [2, 16, 18]. The associated Hamiltonian for this system is given by the following:

$$H = \frac{1}{2}\sum_{n=1}^{N} v_n^2 + \frac{1}{2}\sum_{n \neq m}^{N} \frac{|l_{mn}|^2}{(x_m - x_n)^2}. \quad (2)$$

As in Eq. (1), $H_0$ is given by the first term in the expression and $ZH_b$ by the latter. The level dynamics of this system is governed by the following closed set of ordinary differential equations:

$$\frac{dx_m}{d\lambda} = v_m,$$

$$\frac{dv_m}{d\lambda} = 2\sum_{m \neq n} \frac{|l_{mn}|^2}{(x_m - x_n)^3}, \quad (3)$$

$$\frac{dl_{mn}}{d\lambda} = \sum_{k \neq m, n} l_{mk} l_{kn} \left( \frac{1}{(x_m - x_k)^2} - \frac{1}{(x_k - x_n)^2} \right),$$

where $x_m(\lambda) = E_m(\lambda) = \langle m | H | m \rangle$, denoting the instantaneous eigenvalues of the system, $v_m(\lambda) = \langle m | ZH_b | m \rangle$ and $l_{mn}(\lambda) = (E_m(\lambda) - E_n(\lambda)) \langle m | ZH_b | n \rangle$ satisfying $l_{mn} = -l_{nm}^*$. These represent the "positions", "velocities", and particle-particle repulsion as determined by the "relative angular momenta" [2, 14–21, 22]. These equations have been extended to the stochastic sense, accommodating for noise arising from random fluctuations in the environment affecting the level dynamics of the system [22]. Using the central limit theorem, the sum of these contributions gives a random term in the Hamiltonian with independently Gaussian distributed elements. Using the central limit theorem, noise arises from a number of independent sources; therefore, it is reasonable to assume the sum of its effects is Gaussian. Adding a noise term to the Hamiltonian to represent the system interacting with the environment, the system is transformed from Eq. (1) to

$$H(\lambda(t)) = H_0 + \lambda(t) ZH_b + \delta h(\lambda(t)), \quad (4)$$

where $H_0$ denotes the free Hamiltonian, $ZH_b$ is some perturbation as described in Eq. (1) and $\delta h(\lambda(t))$ denotes a stochastic element representing the influences of external noise [2, 18–21]. To ensure real eigenvalues, the matrix $\delta h$ is Hermitian, hence the diagonal entries are real. This accommodates a range of stochastic systems, for a Brownian model describing a white or colored spectrum. Hence, in the stochastic model, Eq. (2) becomes the following:

$$\dot{x}_m = v_m + \delta h_{mm},$$

$$\dot{v}_m = 2\sum_{m \neq n} \frac{|l_{mn}|^2}{(x_m - x_n)^3} + \frac{l_{mn} \dot{\delta h}_{nm} - \dot{\delta h}_{mn} l_{nm}}{(x_m - x_n)^2},$$

$$\dot{l}_{mn} = \sum_{k \neq m, n} l_{mk} l_{kn} \left( \frac{1}{(x_m - x_k)^2} - \frac{1}{(x_k - x_n)^2} \right) \quad (5)$$

$$+ \frac{(x_m - x_n)(l_{mk} \dot{\delta h}_{km} - \dot{\delta h}_{mk} l_{km})}{(x_m - x_k)(x_n - x_k)}$$

$$+ \dot{\delta h}_{mn}(v_m - v_n) + \frac{l_{mn}(\delta h_{mm} - \delta h_{nn})}{(x_m - x_n)},$$

where the derivative is taken with respect to λ. These equations encode the same level dynamics as the standard Pechukas equations whilst accounting for noise, in the case δh goes to 0, they reduce to the same form, retaining the key feature of an exact mapping of quantum eigenvalue dynamics to a classical gas, independent of any assumptions on the nature of the noise, therefore applicable to a wide range of physical systems [22]. Given the dependence of Eq. (5)

on the derivative of $\delta h$, the noise term must obey a simple stochastic differential equation. An Ornstein–Uhlenbeck process for colored noise was explored in this formalism [22]; however, the model is not restricted in this way and allows for all noise models. Noise provides a source of decoherence in a quantum system.

The evolution of quantum states has been studied through the lens of eigenvalue dynamics in [3], through approximation schemes, adiabatic expansions, time-dependent perturbation theory, and the Magnus series, where it was shown that the latter had the best convergence in short times. In this paper, we move beyond approximations and provide an exact description of quantum state dynamics via the evolution of the eigenvalues. In Sec. 3, this description from eigenvalue dynamics is extended to the evolution of the density matrix via a master equation. This provides insight into the dynamics of occupation numbers and the coherences in their entirety from eigenvalue dynamics, which will prove useful in determining the probability for the system to remain in its initial state. Using this description, one can, for example, determine the effects of avoided level crossings on the system's evolution and the extent to which the noise affects the population of states. Under this description, steady states can be evaluated such that external perturbations do not lead to interferences in the evolution of occupation dynamics. This carries a large number of applications from adiabatic quantum computing to the study of quantum phase transitions and steady states.

## 3. Evolution of the density matrix and coherences

An exact description of the evolution of the quantum states of a many-body open system provides valuable insights into decoherence from quantum phase transitions and, more broadly, into the interactions between the environment and the system. Furthermore, it offers an understanding of coupling between states [6, 7, 17, 23, 24]. However, it remains a monumental challenge due to the amount of information required. This is commonly approached using the master equation given by the stochastic Liouville von Neumann equation [6, 7, 17, 23, 24]:

$$\dot{\rho} = -i[H(t), \rho], \qquad (6)$$

where $\rho$ is Hermitian, denoting the reduced density matrix of the system obtained by tracing out the degrees of freedom of the environment [6, 7, 17, 23, 24], $\dot{\rho}$ refers to the time derivative $\frac{\partial \rho}{\partial t}$, and $H(t)$ denotes a time varying Hamiltonian. The exact form of the Hamiltonian depends on the nature of the system and its coupling to its environment. This may include the influences of the stochastic perturbations resulting in decoherences [6, 7, 17, 23, 24]. Evaluating the dissipation in the system allows for investigating whether the system is driven to a fixed point, the steady state, or a steady state manifold for degenerate fixed points, where diagonal entries of $\dot{\rho} = 0$ [6, 7, 17, 23, 24]. This has direct connotations in AQC, such that the deviation from ground states for a system interacting with its environment can be investigated [23].

Using the Schrödinger equation and the definition of the density matrix combined with the Pechukas–Yukawa formalism, we obtain an analytical expression for the evolution of the density matrix with respect to the level dynamics of the system without simplifying the system to 2 levels, $u$ and $\omega$ and accounting for non-linear evolutions in $\lambda$, given by the following:

$$\dot{\rho}_{uw} = \dot{\lambda} \sum_n \left( \frac{l_{un}}{(x_u - x_n)^2} \rho_{nw} - \rho_{un} \frac{l_{wn}}{(x_w - x_n)^2} \right)$$
$$- i(x_u - x_w)\rho_{uw}, \qquad (7)$$

where $n \neq \omega$ and $n \neq u$. Including the influence of external perturbations, we have the following:

$$\dot{\rho}_{uw} = \dot{\lambda} \sum_n \left( \frac{l_{un}}{(x_u - x_n)^2} + \frac{\delta h_{un}}{x_u - x_n} \right) \rho_{nw}$$
$$- \rho_{un} \left( \frac{l_{nw}}{(x_w - x_n)^2} - \frac{\delta h_{nw}}{x_n - x_w} \right) - i(x_u - x_w)\rho_{uw}. \qquad (8)$$

Separating terms within an $\varepsilon$ neighborhood where $\varepsilon > 0$ and small to consider nearest neighbor interactions

$$\dot{\rho}_{uw} = \dot{\lambda} \sum_{n=\omega-\varepsilon}^{\omega+\varepsilon} \left( \frac{l_{un}}{(x_u - x_n)^2} + \frac{\delta h_{un}}{x_u - x_n} \right) \rho_{nw} - \rho_{un} \left( \frac{l_{nw}}{(x_w - x_n)^2} - \frac{\delta h_{nw}}{x_n - x_w} \right) - i(x_u - x_w)\rho_{uw}$$
$$- \dot{\lambda} \sum_{n=0}^{\omega-\varepsilon} \left( \frac{l_{un}}{(x_u - x_n)^2} + \frac{\delta h_{un}}{x_u - x_n} \right) \rho_{nw} - \rho_{un} \left( \frac{l_{nw}}{(x_w - x_n)^2} - \frac{\delta h_{nw}}{x_n - x_w} \right) \qquad (9)$$
$$- \dot{\lambda} \sum_{n=\omega+\varepsilon}^{N} \left( \frac{l_{un}}{(x_u - x_n)^2} + \frac{\delta h_{un}}{x_u - x_n} \right) \rho_{nw} - \rho_{un} \left( \frac{l_{nw}}{(x_w - x_n)^2} - \frac{\delta h_{nw}}{x_n - x_w} \right).$$

When $\omega = 0$, all terms under the second summation do not exist, and similarly for $\omega = N$, all interactions under the last summation do not exist. This emphasizes the endurance of the ground state being more robust against decoherence.

Considering levels $|u - w| > \varepsilon$, the terms can be disregarded as approximation as nearest neighbor interactions are more dominant. Then, for a general level $\omega$,

$$\dot{\rho}_{uw} = -\dot{\lambda} \sum_{n=\omega-\varepsilon}^{\omega+\varepsilon} \left( \frac{l_{un}}{(x_u - x_n)^2} + \frac{\delta h_{un}}{x_u - x_n} \right) \rho_{nw}$$

$$-\rho_{un}\left( \frac{l_{nw}}{(x_w - x_n)^2} - \frac{\delta h_{nw}}{x_n - x_w} \right) - i(x_u - x_w)\rho_{uw}. \quad (10)$$

The diagonal entries denote the occupation numbers, given by the following:

$$\dot{\rho}_{ww} = -\dot{\lambda}\sum_{n=\omega-\varepsilon}^{\omega+\varepsilon}\left(\frac{l_{wn}\rho_{nw} - \rho_{wn}l_{nw}}{(x_w - x_n)^2} + \frac{\delta h_{wn}\rho_{nw} - \rho_{wn}\delta h_{nw}}{x_w - x_n}\right). \quad (11)$$

These represent the probability of remaining in a state as the system evolves, whereas the off-diagonal entries describe the coherences, superposition of states due to the interactions between the system and the environment, resulting in dephasing the system. This formalism offers valuable insights into the sources of decoherence, attributed to quantum phase transitions and interactions with the environment via noise. Under this description, one could investigate the differences in scaling properties observed between edge and intermediate state transitions, observed in [18, 21]. The interplay between noise and state transitions is generally detrimental to AQC; however, in [22] it was shown that the influences of stochastic perturbations inherently break any hidden symmetries in the Hamiltonian and, thus, split any level crossings in the energy spectrum, suppressing Landau–Zener (LZ) transitions. In Sec. 4, we apply this exact description to the evolution of quantum states in both closed and open systems to observe the influences of noise on the occupation dynamics, such that we can evaluate quantum phase transitions in open systems.

### 4. Two-qubit Ising model

Considering the two-qubit Ising model with the Hamiltonian,

$$H(\lambda(t)) = J\sigma_1^z\sigma_2^z + \lambda Zh_1\sigma_1^x + \lambda Zh_2\sigma_2^x, \quad (12)$$

for the case that $J > 0$ the interaction favors antiferromagnetism, whereas for $J < 0$ it favors ferromagnetism, we take random values for $J$, Gaussian distributed with mean 0 and standard deviation 1, reflecting the different initial conditions. When $J \gg \lambda Zh_1, \lambda Zh_2$ the system is in the ground state. The perturbation matrix is defined by $\lambda ZH_b = \lambda Zh_1\sigma_1^x + \lambda Zh_2\sigma_2^x$, with a large bias, $Z = 10$ and taking $h_1 = 0.1$ and $h_2 = 0.2$. From this, we obtain the values for $x_n$ from the eigenvalues of the system given by $x_n(\lambda) = E_n(\lambda) = \langle n|\mathcal{H}|n\rangle$ the variables for velocity is determined by the following, $v_n(\lambda) = \langle n|Z\mathcal{H}_b|n\rangle$ and relative angular momentum, $l_{mn}$ using its definition that $l_{mn}(\lambda) = (E_m(\lambda) - E_n(\lambda))\langle m|Z\mathcal{H}_b|n\rangle$. $\sigma_j^z$ and $\sigma_j^x$ represent the corresponding Pauli matrices for the $j$th qubit.

The Hamiltonian reads in matrix form

$$\mathcal{H}(\lambda(t)) = J\begin{pmatrix} 1 & 0 & 0 & 0 \\ 0 & -1 & 0 & 0 \\ 0 & 0 & -1 & 0 \\ 0 & 0 & 0 & 1 \end{pmatrix} + \lambda Zh_1\begin{pmatrix} 0 & 0 & 1 & 0 \\ 0 & 0 & 0 & 1 \\ 1 & 0 & 0 & 0 \\ 0 & 1 & 0 & 0 \end{pmatrix}$$

$$+\lambda Zh_2\begin{pmatrix} 0 & 1 & 0 & 0 \\ 1 & 0 & 0 & 0 \\ 0 & 0 & 0 & 1 \\ 0 & 0 & 1 & 0 \end{pmatrix}. \quad (13)$$

The coordinates for $x_n$ are of the form $J + \lambda H_n$. Given that the values for $J$ are Gaussian distributed denoted by, $J \sim \mathcal{N}(\mu, \sigma)$ the values for each $x_n$ are also Gaussian distributed, varying only by a translation by $H_n$ hence, $x_n \sim \mathcal{N}(\mu + \lambda H_n, \sigma) = \mathcal{N}(\lambda H_n, 1) := \tilde{\mathcal{N}}_n$, with the same mean and standard deviation where $H_n = \{-h_1 - h_2, -h_1 + h_2, h_1 - h_2, h_1 + h_2\}$. The values for $v_n$ are deterministic, we define them as $v_n \sim \delta_{H_n}$. We observe that the terms describing $l_{mn}$ determined from its definition are described by the product of the translated Gaussian distributions with the Dirac distributions for $H_n$. The level dynamics for this system are outlined in Fig. 1, where multiple level crossings are observed.

The density matrix $\rho_0$ is initialized with equal probability on a superposition of its eigenstates such that $\rho_0$ is as follows:

$$\mathcal{H}(\lambda(t)) = \frac{1}{4}\begin{pmatrix} 1 & 1 & 1 & 1 \\ 1 & 1 & 1 & 1 \\ 1 & 1 & 1 & 1 \\ 1 & 1 & 1 & 1 \end{pmatrix}. \quad (14)$$

Implementing the master equation defined in Eq. (7), the evolution of the density matrix can be evaluated. Under this formalism, the occupation dynamics are represented in Fig. 2, highlighting the evolution of the states and as the system evolves and the levels interact. The dynamics are then governed entirely by the initial conditions of both the level dynamics and $\rho_0$. To investigate this system further under more generalized initial conditions, a sample of $J$ from $\mathcal{N}(0,1)$ could be used to observe the influence of initial conditions for the eigenvalue dynamics on the evolution of the occupation numbers to determine the statistical properties of the density matrix for this system governed by the initial conditions. Similarly, one could consider a sample of different $\rho_0$ sitting on the Bloch sphere; however, this is beyond the scope of this paper.

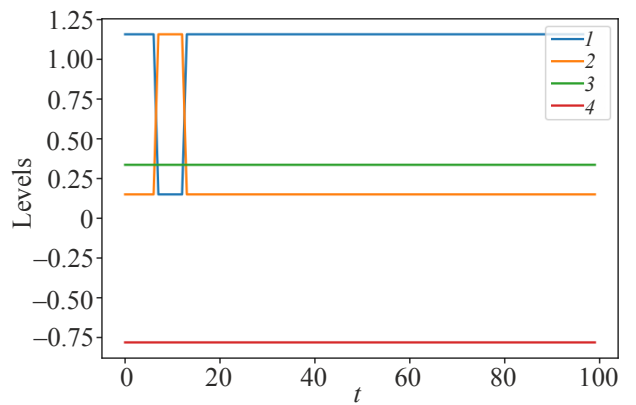

*Fig. 1.* (Color online) The time evolution of the energy levels (*1–4*) of the two-qubit Ising model, with $\lambda$ evolving adiabatically, given by $\lambda = 10^{-3}\log(0.1t)$ for $t \in [0,100]$. Level dynamics are encoded in the initial conditions, governed by $\lambda$ being a function of time. Multiple level crossings are observed between the different levels as they evolve. We note that the levels are seen to be moving away from each other as time evolves.

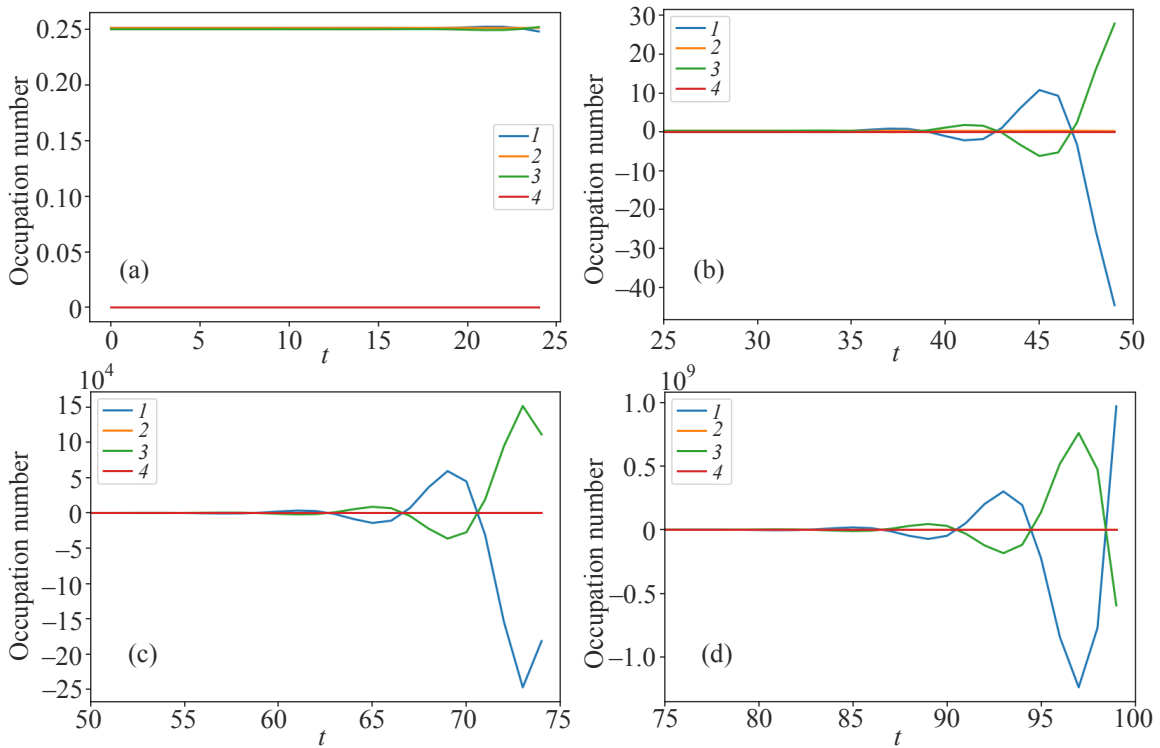

*Fig. 2.* (Color online) The occupation dynamics corresponding to the evolution of the quantum states (*1–4*) of the adiabatically evolving two-qubit Ising model with $\lambda = 10^{-3}\log(0.1t)$ for $t \in [0,100]$. The system is initialized in a superposition of the eigenstates associated with the free Hamiltonian with equal probability amplitudes. It is observed how the occupation numbers evolve with the level dynamics as $\lambda(t)$ grows. The plots are taken in segments of 25-second intervals to accommodate the changes in scale as occupation numbers evolve.

As detailed in Eq. (A7), this formalism is capable of describing the evolution of state dynamics in open systems where the system interacts with its environment. Considering the two-qubit Ising model interacting with its environment. Under the assumptions of the central limit theorem, which states that if you take sufficiently large samples from a population, the samples means will be normally distributed, even if the population isn't normally distributed, then taking $\delta h$ to represent a Weiner process $W_{\lambda(t)}$ where its derivative is a normal distribution denoted by $dW_{\lambda(t)} = \sqrt{(d\lambda(t))}\mathcal{N}(0,1)$. With the additional condition that the noise matrix is Hermitian, the following dynamics is observed in Fig. 3.

Level crossings occur throughout the evolution of the eigenvalues, and using the stochastic master equation defined in Eq. (A7), the evolution of the quantum states can be evaluated exactly. The evolution of the occupation numbers is observed in Fig. 4.

Again, the dynamics are governed entirely by the initial conditions of both the level dynamics and $\rho_0$, however, the stochastic variability in the levels results in different level dynamics despite the same initial conditions. In this instance, divergences occurred sooner in the occupation numbers as neighboring levels approached a level crossing, as observed in the level dynamics in Fig. 3, resulting in the occupation numbers diverging further apart.

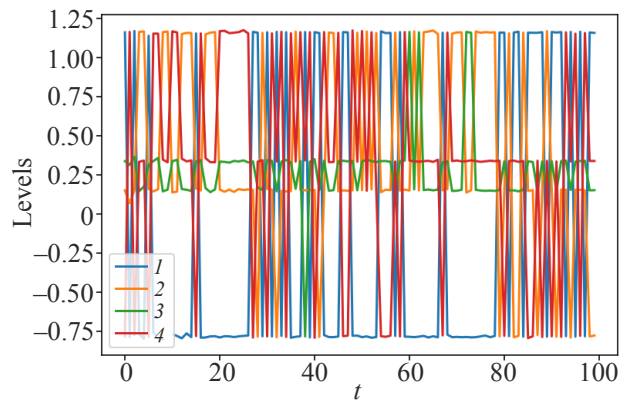

*Fig. 3.* (Color online) The evolution of the energy levels (*1–4*) of the two-qubit Ising model under the influence of noise. Again, the level dynamics are encoded in the initial conditions, governed by $\lambda$ adiabatically evolving as a function of time given by $\lambda = 10^{-3}\log(0.1t)$, for $t \in [0,100]$. Multiple level crossings are observed between the different levels as they evolve.

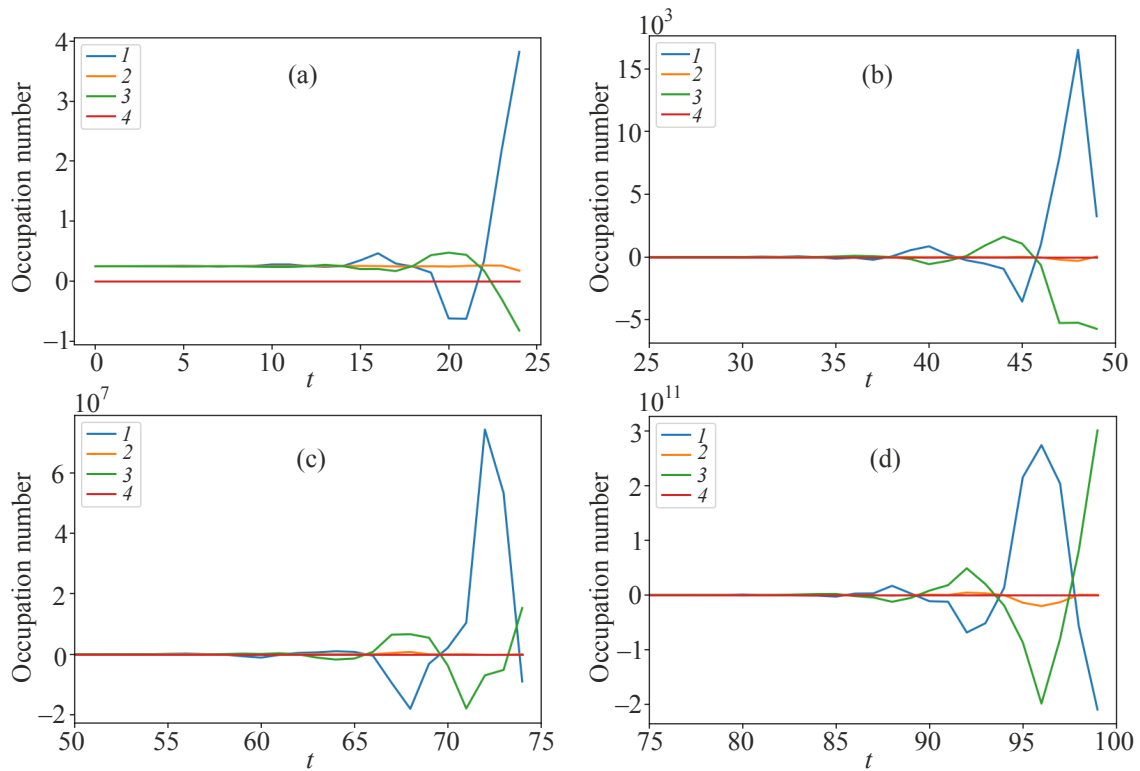

*Fig. 4.* (Color online) The occupation dynamics corresponding to the adiabatic evolution of the quantum states (*1–4*) of the two-qubit Ising model under the influence of a Wiener process. $\lambda(t)$ is given by, $\lambda = 10^{-3} \log(0.1t)$ for $t \in [0,100]$. The system is initialized in a superposition of the eigenstates associated with the free Hamiltonian with equal probability amplitudes. It is observed how the occupation numbers evolve with the level dynamics as $\lambda(t)$ grows. The plots are taken in segments of 25 s intervals to accommodate the changes in scale as occupation numbers evolve.

As demonstrated, this formalism provides an exact description of the quantum state dynamics via eigenvalue dynamics. This plays a crucial role in the understanding of non-equilibrium quantum phase transitions. Using the Pechukas–Yukawa formalism, it would be possible to evaluate stochastic QPT and investigate steady states and how their properties could be harnessed for the novel approach to AQC. To consider QPT under dissipative influences, it may be possible to investigate the manifestation of critical behavior in steady states rather than ground states. Adiabatic systems reaching a steady state remain in the steady state. This has the advantage that when the system deviates from the steady state because of stochastic influences, it returns to the steady state. In the realms of AQC, decoherence could be monitored such that the system approaches a desired steady state manifold under controlled dissipative systems. Furthermore, our understanding of non-equilibrium QPT is limited in contrast to equilibrium QPT. One reason for this is that simulating the evolution of a master equation proves more difficult than Hamiltonian dynamics due to the sheer amount of information required for the density matrix, in contrast to the wavefunction. The influ-ence of noise on the evolution of the density matrix and hence the quantum system is investigated using eigenvalue dynamics via the Pechukas–Yukawa formalism, building on previous works in [3], such that this scheme is without approximation, accounting for interactions between all levels as well as allowing for the description of open quantum systems.

## 5. Discussion and conclusions

This paper presents the stochastic Liouville von Neumann master equation in the Pechukas–Yukawa framework. Hence, state dynamics for an open perturbed quantum system can be described analytically through the lens of level dynamics governed by its initial conditions. This goes beyond previous works where perturbation approximations have been adopted [3, 4, 17] and offers insights into the interplay between the level dynamics of a system and the impacts it has on state occupation numbers and coherences between interacting states, crucial to the development of AQC. Furthermore, this description allows for an improved understanding of the influences of noise due to coupling from the environment on the state dynamics. Given the versatility of this framework, there is potential to have a significant impact across a broad range of disciplines. Whilst being less computationally exhausting, sources of decoherence can be better evaluated. The main sources of decoherence arise due to interactions from the environment, resulting in noise in the system or variation in occupation numbers due to quantum phase transitions. In this framework, both these sources can be investigated through the level dynamics. This description provides a richer understanding of the mechanisms associated with LZ transitions and hence quantum phase transitions. The standard LZ model deals only with the 2 interacting levels. Extending to the multistate problem could yield more interesting physics analytics. The Pechukas–Yukawa model concerns an interacting system of $N$ entangled levels. It is highly equipped to consider interacting systems with entangled states. In further works, it would be useful to consider the detailed analytics of multiple-level interactions and their influence on each other's dynamics to better understand the occupation dynamics and evolution of coherences independent of the conditions imposed on the system by the Landau–Zener model. This is valuable in studying the dynamics of adiabatic systems, crucial to the development of adiabatic quantum computers. Additionally, under this description, steady states and steady state manifolds can be studied via eigenvalue dynamics. This may prove valuable in evaluating steady state trajectories in AQC as the system graduates from its ground state. Moreover, identifying steady states in open quantum systems and the speed of approach can be investigated.

Another area of research, in addition to the investigation of decoherence, would be to explore the chaotic behavior in large quantum systems and shed light on state interactions [17]. Beyond AQC, this framework may prove useful in exploring optimization solutions involved in quantum reservoir computing and studying chaotic time series.

## Acknowledgments

I am grateful for the fruitful discussions with Dr. A. Zagoskin on this topic.

## Appendix A: Derivation of occupation dynamics via Pechukas–Yukawa formalism

Starting with a general perturbed Hamiltonian in Eq. (1) and using the stochastic Liouville von Neumann equation $\frac{\partial \rho}{\partial t} = \frac{-i}{\hbar}[H,\rho]$, we arrive at the following with $\hbar = 1$,

$$\frac{\partial \rho}{\partial t} = -i[H_0 + \lambda(t)ZH_b + \delta h(\lambda(t)), \rho]. \quad (A1)$$

We take $\delta h(\lambda t))$ to be white noise such that expectation is 0: $<\delta h(\lambda(t))> = 0$ and covariance a $\delta$ distribution: $<\delta h(\lambda(t)), \delta h(\lambda(t))> = K\delta(t-t')$. Using the definition of the density matrix, $\rho = \sum_{m,n} \rho_{mn} |m><n|$ the RHS is transformed to the following:

$$-i[H,\rho]$$
$$= \sum_{m,n} |\dot{m}> \rho_{mn} <n| + |m> \dot{\rho}_{mn} <n| + |m> \rho_{mn} <\dot{n}|. \quad (A2)$$

Applying the Hamiltonian to the states, the LHS becomes $-i\sum_{m,n}(x_m - x_n)\rho_{mn}$. Using eigenvalue dynamics such that, $H|m> = x_m|m>$ and applying an orthonormal basis state $<n|$ from the right, we obtain the following:

$$<n|\dot{H}|m> = (x_m - x_n)<n|\dot{m}>. \quad (A3)$$

Substituting the general Hamiltonian description in Eq. (1), combined with the Pechukas–Yukawa Hamiltonian Eq. (2), we obtain the following:

$$<n|\dot{m}> = -\dot{\lambda}\left(\frac{l_{mn}}{(x_n - x_m)^2} - \frac{\delta h_{mn}}{(x_n - x_m)}\right). \quad (A4)$$

Applying to Eq. (A2) states $<u|$ from the left and $|w>$ from the right,

$$<u|-i\sum_{m,n}(x_m - x_n)\rho_{mn}|w> = \sum_{m,n} <u|\dot{m}>\rho_{mn}<n|w> + <u|m>\dot{\rho}_{mn}<n|w> + <u|m>\rho_{mn}<\dot{n}|w>. \quad (A5)$$

Substituting the relation in Eq. (A4)

$$-i(x_u - x_w)\rho_{uw} = -\dot{\lambda}\sum_m \left(\frac{l_{um}}{(x_u - x_m)^2} - \frac{\delta h_{um}}{(x_u - x_m)}\right)\rho_{mw} + \dot{\rho}_{mn} + \dot{\lambda}\sum_n \rho_{un}\left(\frac{l_{wn}}{(x_n - x_w)^2} - \frac{\delta h_{nw}}{(x_n - x_w)}\right)\rho_{mw}. \quad (A6)$$

Both $m$, $n$ are dummy variables in the summations, as such, we take the terms under the same sum

$$\dot{\rho}_{uw} = \dot{\lambda}\sum_n \left(\frac{l_{un}}{(x_u - x_n)^2} + \frac{\delta h_{un}}{x_u - x_n}\right)\rho_{nw}$$

$$-\rho_{un}\left(\frac{l_{wn}}{(x_w - x_n)^2} - \frac{\delta h_{nw}}{x_n - x_w}\right) - i(x_u - x_w)\rho_{uw}. \quad (A7)$$

Eq. (A7) is the stochastic Liouville master equation, representing both the evolution of occupation of states and coherences without compromise.

———————————————

Еволюція основних рівнянь Ліувілля фон Неймана
в межах моделі Печукаса–Юкави

Mumnuna A. Qureshi

Представлено новий формалізм для нерівноважної динаміки матриці густини, здатний описувати сильно заплутані багаточастинкові взаємодії. Еволюцію квантових станів оцінюють за допомогою динаміки власних значень загальної гамільтонової системи, збуреної параметрично еволюціонуючою змінною λ(*t*), яка несе залежність від часу. Це досягається за допомогою відображення Печукаса–Юкави еволюції енергетичних рівнів, що визначаються їхніми початковими умовами, на узагальненій моделі Калоджеро–Сазерленда одновимірного класичного газу. Таким чином, квантові системи можна точно описати в повному обсязі за допомогою динаміки власних значень. Згідно з цим описом, надано покращене розуміння взаємозв'язку між нерівноважними квантовими фазовими переходами та декогеренцією, що має значний вплив на широкий спектр застосувань.

Ключові слова: сильно заплутані багаточастинкові взаємодії, квантові стани, відображення Печукаса–Юкави, модель Калоджеро–Сазерленда, нерівноважні квантові фазові переходи.